\documentclass{elsarticle}

\usepackage{amssymb}

\usepackage[figuresright]{rotating}

\usepackage{color}
\definecolor{myred}{rgb}{0.92,0.0,0.}

\begin{document}

\begin{frontmatter}




\title{Anomaly in the $K^0_S\,\Sigma^+$ photoproduction cross section
  off the proton at the $K^*$ threshold}


\author[label1]{R. Ewald\fnref{label2}}
\author[label3,label4]{A.V.~Anisovich} 
\author[label1]{B. Bantes}
\author[label3]{O. Bartholomy}
\author[label3,label4]{D. Bayadilov}
\author[label3]{R. Beck}
\author[label4]{Y.A. Beloglazov}
\author[label3]{K.-T.~Brinkmann}
\author[label3,label5]{V.~Crede} 
\author[label1]{H. Dutz}
\author[label1]{D.~Elsner}
\author[label1]{K.~Fornet-Ponse}
\author[label1]{F.~Frommberger}
\author[label3]{Ch.~Funke}
\author[label4]{A.~B.~Gridnev}
\author[label3]{E.~Gutz}
\author[label1]{W.~Hillert}
\author[label1]{J.~Hannappel}
\author[label3]{P.~Hoffmeister}
\author[label6]{I.~Jaegle}
\author[label1]{O.~Jahn}
\author[label1]{T.~Jude}
\author[label3]{J.~Junkersfeld}
\author[label3]{H.~Kalinowsky}
\author[label1]{S.~Kammer\fnref{label2}}
\author[label1]{V.~Kleber\fnref{label8}}
\author[label1]{Frank Klein}
\author[label1]{Friedrich~Klein}
\author[label3]{E.~Klempt}
\author[label6]{B. Krusche}
\author[label3]{M.~Lang}
\author[KVI]{H.~L\"ohner}
\author[label4]{I.V.~Lopatin}
\author[label1]{D.~Menze}
\author[label6]{T.~Mertens}
\author[KVI]{J.G.~Messchendorp}
\author[label7]{V.~Metag}
\author[label7]{M.~Nanova}
\author[label3,label4]{V.A.~Nikonov}
\author[label3,label4]{D.~Novinski}
\author[label7]{R.~Novotny}
\author[label1]{M.~Ostrick\fnref{label9}}
\author[label7]{L.~Pant\fnref{label10}}
\author[label3]{H.~van Pee}
\author[label7]{A.~Roy\fnref{label11}}
\author[label3,label4]{A.V.~Sarantsev}
\author[label7]{S.~Schadmand\fnref{label12}}
\author[label3]{C.~Schmidt}
\author[label1]{H.~Schmieden\corref{cor1}}
\ead{schmieden@physik.uni-bonn.de}
\cortext[cor1]{corresponding author}
\author[label1]{B.~Schoch}
\author[KVI]{S.~Shende}
\author[label3]{V. Sokhoyan}
\author[label1]{A.~S{\"u}le}
\author[label4]{V.V.~Sumachev}
\author[label3]{T.~Szczepanek}
\author[label3]{U.~Thoma}
\author[label7]{D.~Trnka}
\author[label7]{R.~Varma\fnref{label11}}
\author[label3]{D.~Walther}
\author[label3]{Ch.~Wendel} 
\fntext[label2]{now at DLR, Cologne, Germany}
\fntext[label8]{now at German Research School for Simulation Sciences, J{\"u}lich, Germany}
\fntext[label9]{Present address: Institut f\"{u}r Kernphysik, Universit\"{a}t
Mainz, Germany}
\fntext[label10]{On leave from Nucl. Phys. Division, BARC, Mumbai, India}
\fntext[label11]{On leave from Department of Physics, IIT, Mumbai, India}
\fntext[label12]{Present address: Institut f\"{u}r Kernphysik and J\"{u}lich
Center for Hadron Physics, Forschungszentrum J\"{u}lich, Germany}





\address[label1]{Physikalisches Institut, Rheinische
  Friedrich-Wilhelms-Universit{\" a}t Bonn, Germany}
\address[label3]{Helmholtz-Institut f{\"u}r Strahlen- u. Kernphysik, Rheinische
  Friedrich-Wilhelms-Universit{\" a}t Bonn, Germany}
\address[label4]{Petersburg Nuclear Physics Institute, Gatchina, Russia}
\address[label5]{Department of Physics, Florida State University, Tallahassee,
  USA}
\address[label6]{Department of Physics and Astronomy, University of Basel,
  Switzerland}
\address[KVI]{Kernfysisch Versneller Instituut, Groningen, The Netherlands}
\address[label7]{II. Physikalisches Institut, Universit\"at Gie{\ss}en,
  Germany\\[.2cm] The CBELSA/TAPS Collaboration}

\begin{abstract}
The $\gamma + p \rightarrow K^0 + \Sigma^+$ photoproduction reaction is
investigated in the energy region from threshold to $E_\gamma = 2250$\,MeV.
The differential cross section exhibits increasing forward-peaking with 
energy, but only up to the $K^*$ threshold. 
Beyond, it suddenly returns to a flat distribution
with the forward cross section dropping by a factor of four.
In the total cross section a pronounced structure is observed between the 
$K^*\Lambda$  and $K^*\Sigma$ thresholds.
It is speculated whether this signals the turnover of the reaction mechanism 
from $t$-channel exchange below the $K^*$ production threshold to an
$s$-channel mechanism associated with the formation
of a dynamically generated $K^*$-hyperon intermediate state.
\end{abstract}


\end{frontmatter}


\section{Introduction}
\label{sec:Introduction}

The CBELSA/TAPS experiment at the Electron Stretcher Accelerator ELSA
\cite{Hillert06} of Bonn
University is investigating the structure of the nucleon at low energies.
The excitation spectrum provides a fingerprint of the intra-baryonic 
interaction dynamics in the non-perturbative regime of QCD.
Lattice calculations made impressive progress over recent years and
provide ab initio calculations with almost realistic quark masses 
of the spectrum of ground state baryon masses \cite{Duerr08}.
However, understanding excited states is still a challenge using lattice QCD, 
and so quark models are often used to describe the nucleon excitation 
spectrum.
In a simple though extreme view the baryon structure is determined by a strong
interaction potential which is mutually generated by the constituent quarks.
While corresponding models \cite{HK83,CR00,LMP01}
are successful in reproducing crucial parameters of 
low mass states, e.g. magnetic moments and electromagnetic couplings, 
important aspects of the excitation spectrum still remain dubious:
(i) An essential fraction of higher mass states expected within the 
$SU(6) \times O(3)$ quark models has not yet been observed.
It is an open issue whether this reflects deficits of the models 
or of incomplete experiments.
(ii) Even basic features of low-lying states are difficult to understand in 
genuine 3-quark models, e.g.
the parity ordering of the lowest lying nucleon excitations
$N(1440)P_{11}$ (positive parity) and $N(1535)S_{11}$ (negative parity),
which naturally would be reversed in any three dimensional static 
potential.
This is a point of controversy also in lattice QCD calculations
\cite{Mathur05,Lin11}.
Further prominent problems are the unusually large decay into the $\eta$ 
channel of the $N(1535)S_{11}$ compared to both, the nearby angular momentum 
partner $N(1520)D_{13}$ and the $N(1650)S_{11}$ with even the same quantum 
numbers, 
or the large 115\,MeV mass gap within the angular-momentum doublet 
$\Lambda(1405)S_{01}$ and $\Lambda(1520)D_{03}$.
These facts were speculated to signal other than 3-quark dynamics 
to affect the observed resonance structure.
Associated with the spontaneous breaking of chiral symmetry, 
some models indeed assign an essential role for baryon dynamics 
to meson fields \cite{GR96,MG84}
and meson-baryon interactions 
\cite{Dalitz,SW88,KWW97,Inoue2001,Hyodo2002,G-RLN04,LK04,MRR04,Borasoy07}. 
In particular in the vicinity of thresholds, the formation of (unstable)
hadronic molecules in the sense of states which are dynamically generated 
by baryon-meson interaction is expected.
They should show up, at least, as strong baryon-meson Fock components. 
For some low-lying states meson cloud effects 
seem to be indicated by meson electroproduction experiments at 
small momentum transfers \cite{Aznauryan09},
but a direct experimental observation of molecular components 
is still missing.

\begin{figure}
  \begin{center}
   \includegraphics[width=0.9\textwidth]{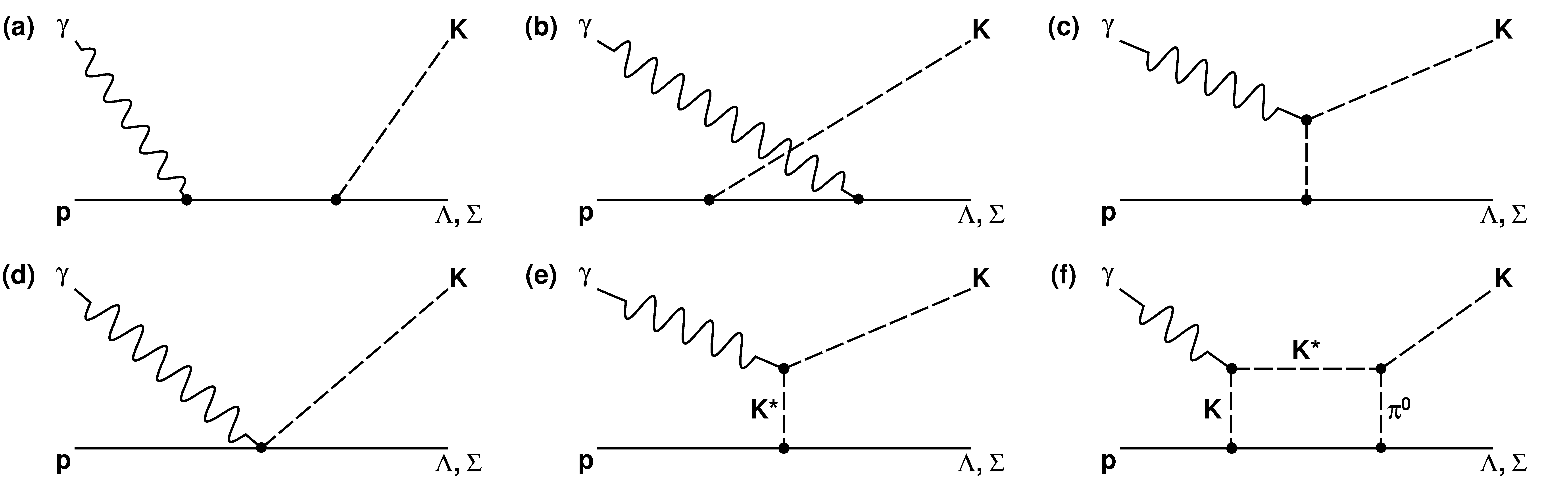}
  \end{center}
  \caption{\footnotesize %
Diagrams contributing to charged and uncharged kaon photoproduction.
The Born diagrams are shown in (a) - (d). Non-strange resonances may
contribute as intermediate states in the $s$-channel (a) and 
$u$-channel (b). The $t$-channel meson exchange (c) and the seagull term
(d) are proportional to the meson charge, hence contribute only to
the charged kaon channel. Vector meson exchange (e) is allowed also
for $K^0$ photoproduction. Diagram (f) visualizes subthreshold $K^*$
production with subsequent coupling into the kaon channel through
$\pi^0$ rescattering, as explained in the text.   
    }
  \label{fig:diagrams}
\end{figure}

Meson photoproduction provides a sensitive tool to investigate these
issues.
In the experiment presented here the reaction
$\gamma \, p \,\rightarrow\, K^0 \, \Sigma^+$ was studied
from threshold ($E_\gamma = 1047.6$\,MeV) to
a photon energy of $E_\gamma = 2250$\,MeV, 
i.e. across the $K^*$ production threshold at 
$E_\gamma = 1678.2$\,MeV for the $K^{*+}\Lambda$ final state, 
and at $E_\gamma = 1848.1$\,MeV for $K^{*0}\Sigma^+$.
Compared to charged $K^+$ photoproduction,
which has been extensively studied during recent years \cite{Bradford07}, 
the neutral $K^0$ channel has tended to be sidelined.
This appears entirely unjustified, though.
To study $s$-channel resonance excitations (Figure\,\ref{fig:diagrams}(a)),
the photoproduction of neutral kaons offers some advantages 
over charged ones,
because the photons cannot directly couple
to the (vanishing) charge of the meson.
Hence, the $t$-channel diagram (c) in Figure \ref{fig:diagrams} 
does not contribute to the production process.
Since this may become dominant in charged kaon photoproduction,
the neutral channel provides a cleaner probe for $s$-channel
excitations.
However, $t$-channel processes are not entirely suppressed in
$K^0$ photoproduction. 
The photon coupling to the $K^0$-$K^{0*}$ vertex remains non-zero
and renders a $t$-channel exchange of a $K^*$ meson possible
as is visualised in Figure\,\ref{fig:diagrams}(e).
This opens the opportunity to get a hand on explicit meson-baryon dynamics:
Should $K^*$-hyperon dynamics play a significant role, 
then $K^*$ production via diagrams of the type 
Figure\,\ref{fig:diagrams}(f)
may yield $K^0$ photoproduction markedly different above and below
the $K^{0*}$ threshold, unmasked by the strong charge-dominated 
t-exchange in $K^+$ production.   
These considerations provided the motivation for our study of the 
$\gamma \, p \, \rightarrow K^0 \, \Sigma^+$ photoproduction reaction
presented here.
Previous data of  Crystal Barrel \cite{Castelijns08} and 
SAPHIR \cite{Lawall05} agree rather well in general,
however differences show up just in the energy region 
of the $K^*$ threshold, prohibiting any clear conclusions on 
$K^*$-hyperon dynamics.
The goal of the present experiment was to improve this unsatisfactory 
situation.

\section{Experiment}
\label{sec:Experiment}

The experiment was performed with the combined 
Crystal Barrel \cite{Aker92} 
and TAPS \cite{Novotny91,Gabler94}
detector system at the tagged photon beam of ELSA 
using an electron beam of $E_0 = 3.2$ GeV.
Bremsstrahlung was produced from a $500\,\mu$m thick diamond crystal.
Coherent bremsstrahlung peaks which carry linear polarisation were 
subsequently set at 1305, 1515 and 1814 MeV.
For the present analysis the azimuthal distributions were summed over, 
however, such that the effective photon linear polarisation was zero.   
The bremsstrahlung electrons were momentum analysed in
a magnetic dipole (tagging-) spectrometer. 
The spectrometer's electron detector covered a photon energy range of
$E_\gamma = 0.18$--$0.92 E_0$.
14 slightly overlapping scintillator bars provided the tagger timing, 
whereas the energy was determined by a 480 channel 
scintillating fibre detector at low photon energies (i.e. high rates),
complemented by a MWPC at high energies (low rates).
An energy resolution between 10 and 25\,MeV was achieved, depending on the
energy of the electron incident on the tagging detector.
The tagging system was run at electron rates up to $10^7$\,Hz.
The absolute photon flux was determined from the tagged electron
spectrum in combination with a fast total absorbing PbWO$_4$ detector
to measure the energy dependent tagging efficiency.

The photon beam hit a $5.3$ cm long 
liquid hydrogen target with 80\,$\mu$m Kapton windows.
A three layer scintillating fibre detector (ISFD) \cite{Suft05} 
surrounded the target within the polar angular range from 
15 to 165 degrees. 
It determined a point-coordinate for charged particles.
Both, charged particles and photons, were detected in the 
{Crystal Barrel} detector.
The 1290 CsI(Tl) crystals of 16 radiation lengths were arranged
cylindrically around the target in 23 rings,
covering a polar angular range of 30 -- 168 degrees, 
and read out by photo diodes attached to a wavelength shifter.
For photons an energy resolution of 
$\sigma_{E}/E 
= 2.5\,\%/ \sqrt[4]{E/\textrm{GeV}}$
and an angular resolution of $\sigma_\textrm{angle} \simeq 1.1$\,degree 
was obtained.

The $5.8$ -- 30 degree forward cone was covered by the {TAPS} detector, 
set up in one hexagonally shaped wall of 528 BaF$_2$ modules
at a distance of $118.7$\,cm from the target.
For photons between 45 and 790 MeV an energy resolution of
$\sigma_{E}/E
= \left(0.59/\sqrt{E/\textrm{GeV}}+1.9\right)\%$ was achieved.
The position of photon incidence could be resolved within
20\,mm.
The {TAPS} detectors were individually equipped with photomultiplier readouts.
Each {TAPS} module had a 5\,mm plastic scintillator in front of it
to measure the energy loss signal of charged particles.
The first level trigger was derived from {TAPS}.
For this purpose the detector was subdivided into 8 sectors of crystals, 
each provided with two discriminator thresholds. 
A TAPS-high trigger corresponds to at least one group above the high
threshold, whereas TAPS-low requires at least two sectors above the low
threshold.
The thresholds were chosen as low as possible while keeping the overall 
TAPS trigger rate at a tolerable level of about 100 kHz. 
A cluster finder algorithm (FAst Cluster Encoder -- FACE) 
for the {Crystal Barrel} provided a second level trigger. 
Here a minimum of either one or two clusters could be demanded.
The detector system was read out upon one of the two global trigger conditions:
(1) TAPS-low without FACE requirement or (2) TAPS-high along with FACE.
In both cases, the coincidence with the tagging spectrometer was 
additionally required.

With the TAPS detector a 1-sigma width of 390\,ps was achieved 
in the time calibration for coincident photon hits in different detector 
modules and of 690\,ps relative to the photon tagger.
Starting with the energy loss induced by cosmics to set the individual
high voltages, 
the energy calibration for both the Crystal Barrel and TAPS detectors
was performed iteratively by an offline gain adjustment to minimise the
width of the $\pi^0$ peak in the 2-photon invariant mass distribution.
In addition, every few hours the Crystal Barrel gains were monitored
by means of a light-pulser system with calibrated attenuators.
The quality of the resulting calorimeter calibrations were then 
cross-checked through the width of the $\eta$ peak in the 6-photon 
invariant mass spectrum.
A 1-sigma width of 39\,MeV was obtained.  
The energy calibration of the tagging system was performed by
deflecting low intensity electron beams of known energy directly 
into the tagging detector plane. 
An absolute accuracy of better than 10\,MeV was achieved.

\section{Event selection and data analysis}
\label{sec:Analysis}

The Crystal-Barrel/TAPS detector setup is optimised for 
multi-photon final states.
Therefore, the $K^0 \Sigma^+$ reaction was studied in the neutral
decay modes $K^0 \rightarrow \pi^0 \pi^0$ (B.R. $31.4\,\%$)
and $\Sigma^+ \rightarrow p \pi^0$ (B.R. $51.6\,\%$),
which in total yields 6 photons along with the proton.
Hence, event topologies with 7 cluster hits in the calorimeters 
were selected.
Charged particles were recognised through their signals in the 
ISFD or the TAPS $\Delta E$ plastic scintillators.
Two charged particle events were discarded.
For one charged particle in the final state the proton assignment was unique.
7 neutral cluster hits were also accepted.
In this case, all combinatorial possibilities for the proton assignment 
were processed.
The resulting combinatorial background was largely reduced through
kinematic requirements in further analysis.
Since the proton detection efficiency was limited,
in particular at low proton momentum,
6 neutral cluster hits were also accepted.   

The tagged photon energy range extended below the $K^0\Sigma^+$
threshold. 
However, in the analysis a photon energy $E_\gamma > 1047.5$\,MeV was 
required to reduce random background.
Furthermore, it was requested that 3 pairs 
of neutral (so-called $\gamma$) hits can be found 
with invariant masses in the range of the $\pi^0$, namely 
$110\,\textrm{MeV} \leq M_{\gamma\gamma} \leq 160\,\textrm{MeV}$.  
The $\eta p \rightarrow p\,6\gamma$ final state provided the most 
significant background.
It was practically eliminated by requiring the invariant mass 
of the three identified $\pi^0$'s to {\em not} lie within the 
range $470\,\textrm{MeV} \leq M_{\pi^0\pi^0} \leq 620\,\textrm{MeV}$.
This corresponds to a 4-sigma wide anti-cut around the $\eta$ mass.
Finally, neutral background events from the electron beamdump
(which was located below the floor in front of the Crystal Barrel calorimeter)
were rejected through their special angular topology.

\begin{figure}
  \begin{center}
   \includegraphics[width=0.80\textwidth]{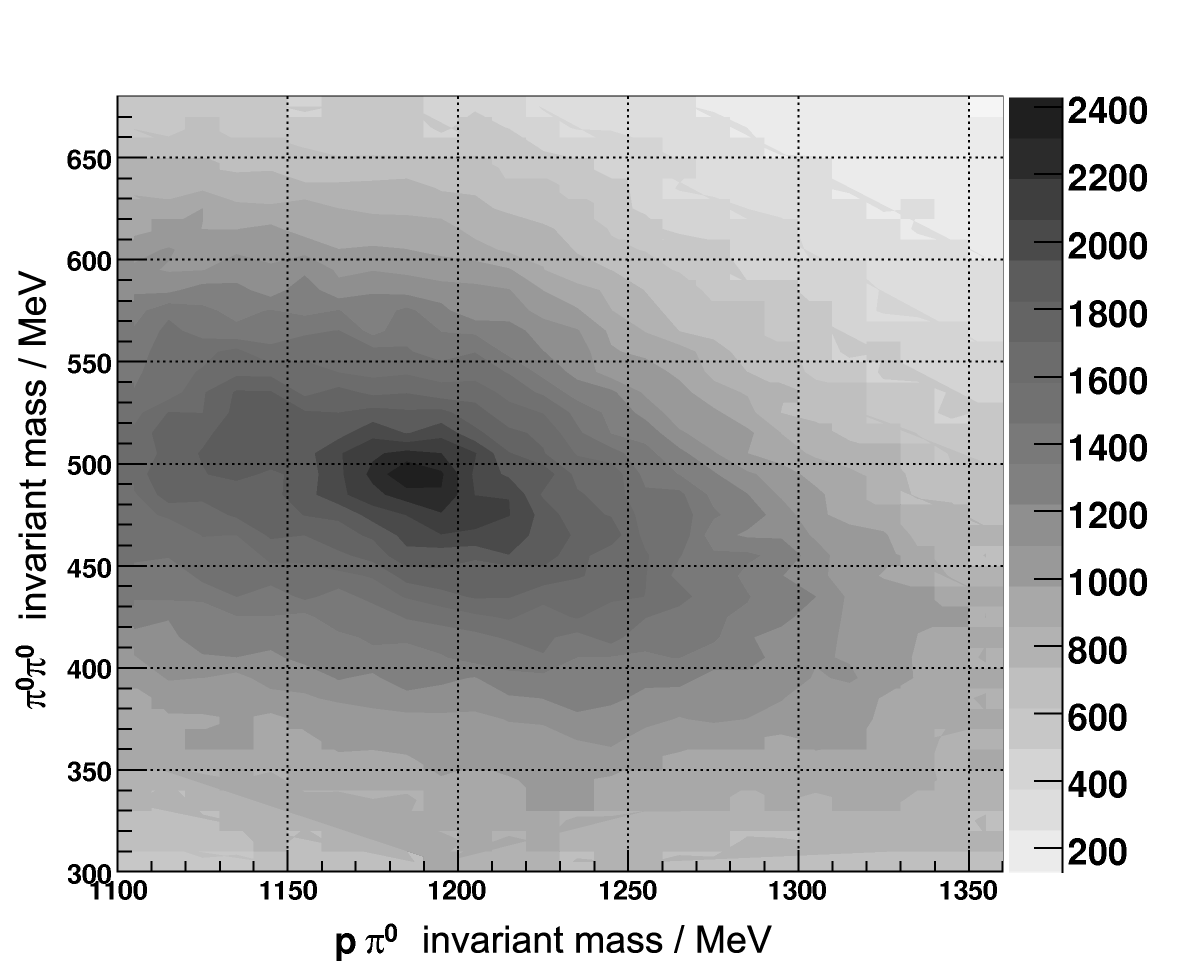}
  \end{center}
  \caption{\footnotesize %
    $\pi^0\pi^0$ against $p\pi^0$ invariant mass distribution after 
    event preselection and kinematic cut, showing a concentration 
    of events in the $K^0\Sigma^+$ final state.
    }
  \label{fig:SigmaVsKaon}
\end{figure}

\begin{figure}
  \begin{center}
   \includegraphics[width=0.6\textwidth]{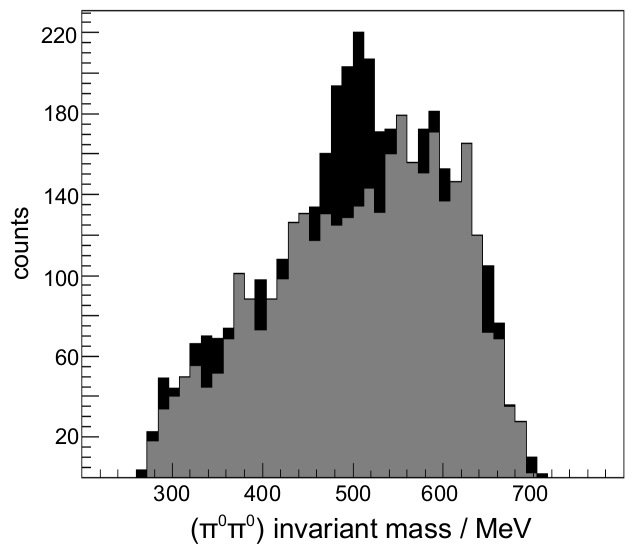}
  \end{center}
  \caption{\footnotesize %
    $\pi^0\pi^0$ invariant mass distribution 
    for the bin 1350 -- 1450\,MeV in photon energy
    and 0 -- $0.33$ in $\cos\Theta^K_\textrm{cms}$
    after a
    cut on the $\Sigma^+$ mass in Figure \,\ref{fig:SigmaVsKaon}:
    $1170\,\textrm{MeV} \leq M_{p\pi^0} \leq 1210\,\textrm{MeV}$.
    The grey area represents the simulated background from
    uncorrelated 3$\pi^0$ photoproduction, scaled to the experimental
    yield outside the signal area (cf. text).
    }
  \label{fig:pi0pi0InvMass}
\end{figure}

Based on this preselection, the remaining events were subjected
to a kinematic fit to the $\gamma \, p \rightarrow p\,3\pi^0$ reaction.
The photon energy was defined by the tagging spectrometer. 
Energy and angle parameters for the fit were provided by the final state 
photon hits. 
The proton candidate did not enter the fit. 
In contrast, its momentum vector was determined through the  
kinematic conditions.
With the four components of the proton momentum to be determined 
and seven conditions (overall energy-momentum and three $\pi^0$ masses)
the fit still was threefold overdetermined.
$\Sigma^+$ and $K^0$ masses were not used as conditions of the 
kinematic fit.
After the kinematic fit an acceptable signal to background ratio
was achieved.
This is demonstrated in Figure\,\ref{fig:SigmaVsKaon} 
where the $2\pi^0$ invariant mass is plotted against 
the $p\pi^0$ invariant mass distribution. 
A culmination of events is clearly visible around 
$M_{\pi^0\pi^0} = 490$ and $M_{p\pi^0} = 1190$\,MeV, corresponding to the 
masses of $K^0$ and $\Sigma^+$.
As an example, Figure\,\ref{fig:pi0pi0InvMass} shows the $\pi^0\pi^0$ 
invariant mass distribution for the bin 1350--1450\,MeV in photon energy
and 0 -- $0.33$ in $\cos\Theta^K_\textrm{cms}$, the kaon center-of-mass angle, 
after a cut on the $\Sigma^+$ mass region in the $p\pi^0$
mass distribution: $1170\,\textrm{MeV} \leq M_{p\pi^0} \leq
1210\,\textrm{MeV}$.
The cut limits were obtained by minimising the related systematic error
induced by background subtraction.
In Monte Carlo simulations it turns out that the far dominating background
is associated with uncorrelated photoproduction of three neutral pions.
The simulated background distribution is shown as the hatched area in
Figure\,\ref{fig:pi0pi0InvMass}.
The spectra agree very well with the experimental distributions in all
bins.
The simulated yield is scaled outside the area of the $K^0$ signal peak.  

The photon energy range was divided into 12 bins of $\pm 50$\,MeV width,
ranging from 1100 to 2200\,MeV. 
Monte Carlo simulations determined the experimental acceptance
individually for each of the three selected event topologies. 
An important benefit of the almost $4\pi$ detection system is the
practically flat acceptance.
\begin{figure}
  \begin{center}
   \includegraphics[width=0.7\textwidth]{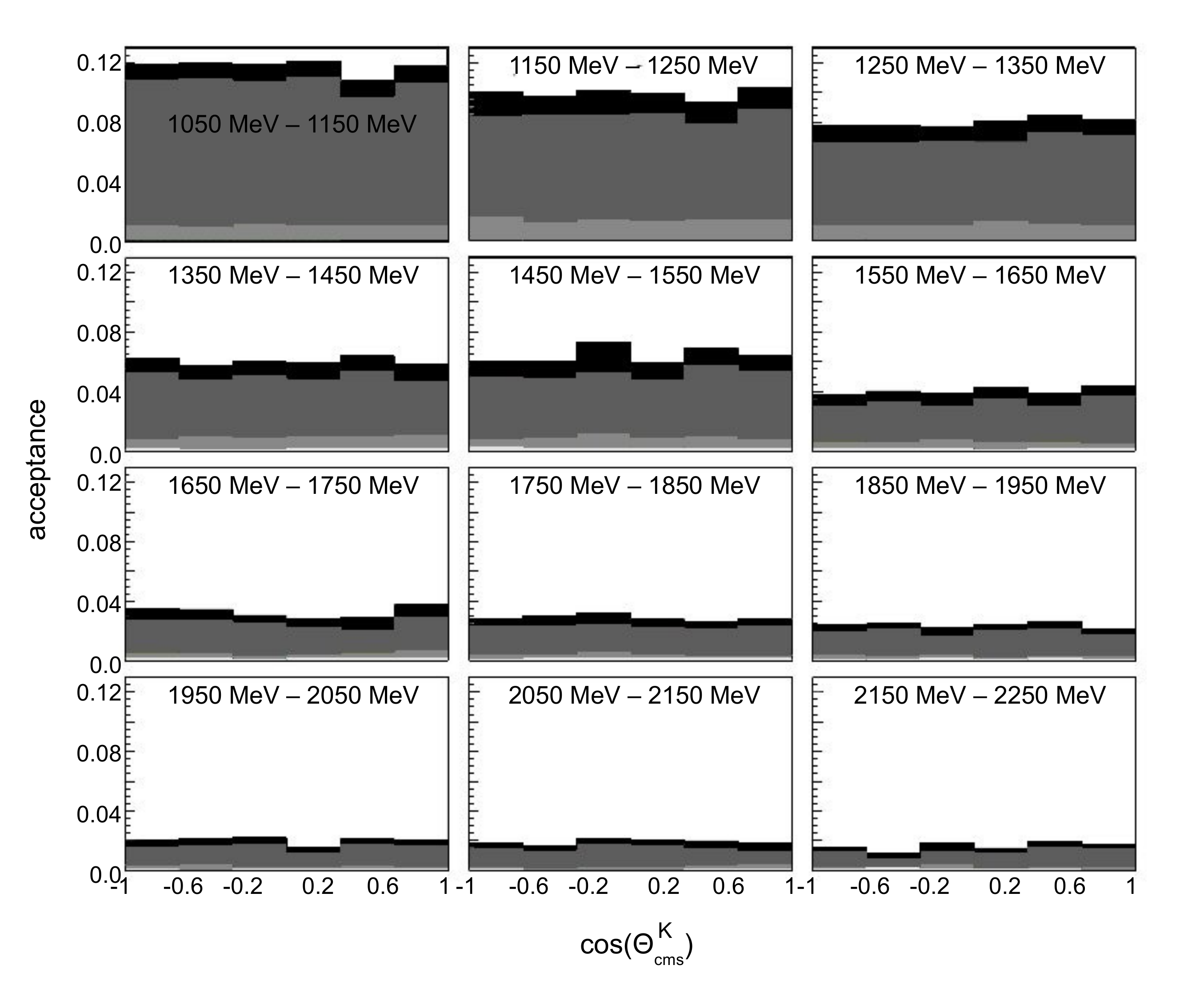}
  \end{center}
  \caption{\footnotesize %
      Simulated acceptance for the $K^0\Sigma^+$ final state. The three
      event topologies are treated separately: 
      6 uncharged and 1 charged hits (grey),
      7 uncharged hits (white), and 6 uncharged hits (light grey).
      The upper histogram (black) represents the total acceptance.
          }
  \label{fig:Accept_Winkel}
\end{figure}
This is visualized in Figure\,\ref{fig:Accept_Winkel}.

\section{Results and Discussion}
\label{sec:Results}

\begin{figure}
  \begin{center}
  \includegraphics[width=0.95\textwidth]{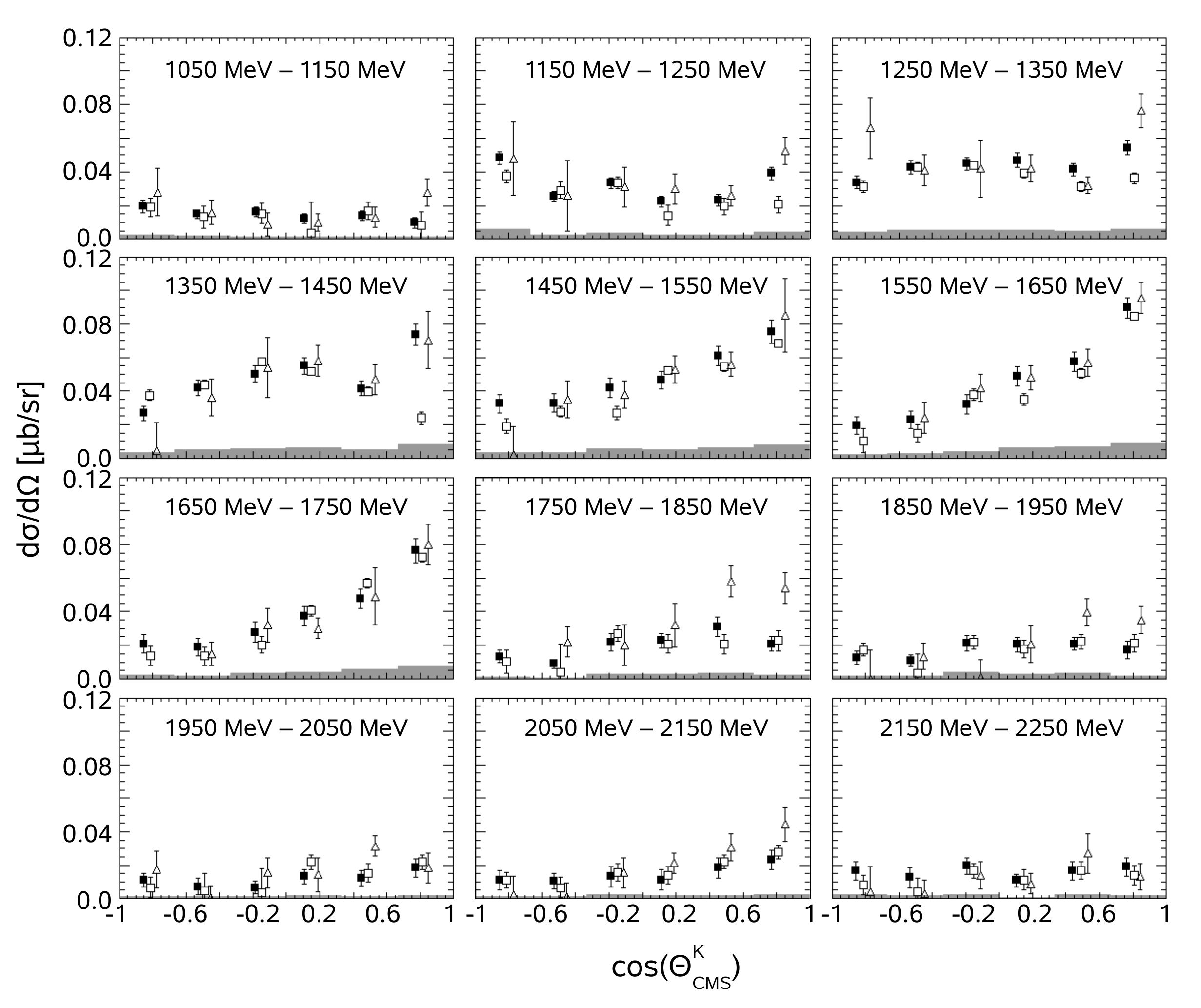}
  \end{center}
  \caption{\footnotesize %
Measured differential cross sections for $K^0\Sigma^+$ photoproduction
as a function of the kaon center-of-mass
angle in $\pm 50$\,MeV wide bins of photon energy from 1100 to 2200 MeV. 
The present results (full squares) are compared to previous measurements
of 
Crystal Barrel (open squares) \cite{Castelijns08} and 
SAPHIR (triangles) \cite{Lawall05}.
The error bars are purely statistical.
An estimate of the systematic uncertainty is given by the bars on the
abscissa (cf. text). 
          }
  \label{fig:DiffWQ_comparison}
\end{figure}

Based on the absolute normalisation of the photon flux provided by
the tagging system, differential cross sections were determined
separately for the 6 and 7 hit topologies.
Both agree very well and were then combined into the full squares 
which are shown in Figure\,\ref{fig:DiffWQ_comparison}.
Associated with the data points are the total statistical errors.
The combined systematic error is indicated by the grey bars on the abscissa.
It has contributions from 
the photon flux ($\simeq 5\,\%$), 
the cuts applied in the analysis ($\simeq 5 - 6\,\%$),
the kinematic fit ($\simeq 5\,\%$) 
and the simulated acceptance ($\simeq 5.6\,\%$). 
The error of the kinematic fit is induced by a cut on the confidence 
level, which was required to exceed $0.1$.  
All errors associated with cuts were estimated through the variation of
the cuts over a wide range. 
Hence, they may be considered as upper error limits.
The most probable errors would be significantly smaller.
Also, the error in the photon flux affects the extracted absolute cross 
sections, but, within a given energy bin, it leaves the form
of the angular distribution unchanged.

In comparison to our new data,
Figure\,\ref{fig:DiffWQ_comparison} also shows the results of 
previous measurements of Crystal Barrel \cite{Castelijns08}
and SAPHIR \cite{Lawall05}.
While the data sets agree relatively well in general, there remain 
significant discrepancies in forward directions and, 
important for the present investigation, in the energy range of 
the $K^*$ threshold ($E_\gamma = 1750$ -- 1850\,MeV bin).
These discrepancies could be resolved by our new data. 
Below the $K^*$ threshold mostly the SAPHIR data is favored.
However, at the $K^*$ threshold and higher energies 
the previous Crystal Barrel data is clearly supported.
The differential cross sections of a CLAS measurement
presented at conferences (not shown in Figure\,\ref{fig:DiffWQ_comparison})
\cite{Carnahan03}, which detected the charged 
$K^0 \rightarrow \pi^+\pi^-$ decay, 
agrees well with our new data within the common detector acceptance.

\begin{figure}
  \begin{center}
\includegraphics[width=0.8\textwidth]{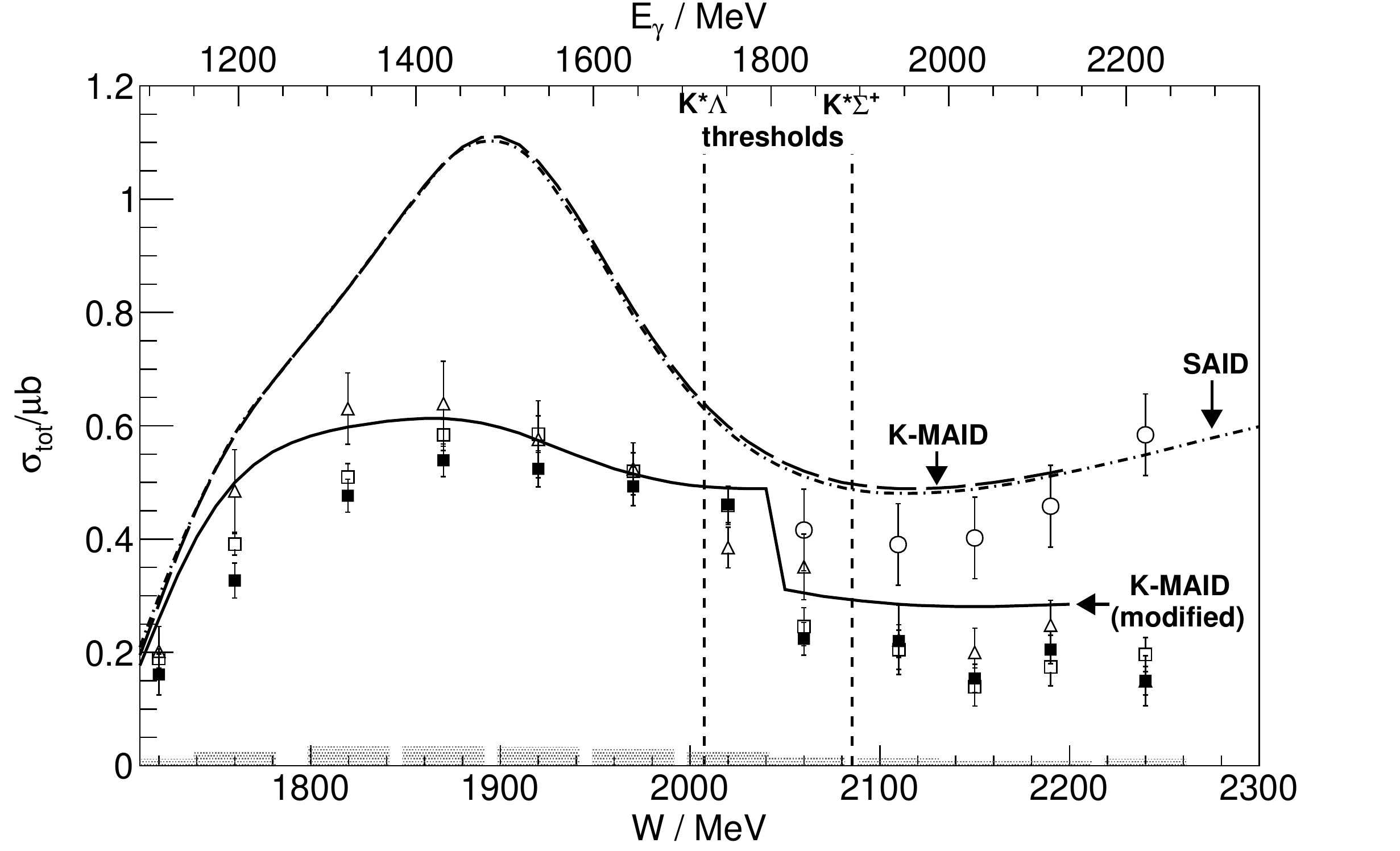}
  \end{center}
  \caption{\footnotesize %
Total cross section for $K^0\Sigma^+$ photoproduction as a function
of the centre-of-mass energy
from the present experiment (full squares)
in comparison to the previous Crystal Barrel (open squares) 
\cite{Castelijns08} and SAPHIR (triangles) \cite{Lawall05} data. 
The vertical lines indicate the 
$K^*\Lambda$ and $K^*\Sigma^+$ thresholds at $W = 2007.4$ and $2085.5$\,MeV,
respectively. 
The SAID parameterisation \cite{SAID} is represented by the dashed-dotted curve.
A K-MAID calculation with standard parameters yields the dashed
curve. The full curve is obtained with the modifications described in the text.
Above the $K^*$ threshold the grey circles represent the sum of the 
$K^0\Sigma^+$ cross section of the present experiment and the $K^{0*}\Sigma^+$
cross section of the work of Nanova et al. \cite{Nanova08}.
The vertical bars on the abscissa again indicate the systematic
error of the present experiment, the errors plotted with the data symbols
are purely statistical.
          }
  \label{fig:total_x-sec}
\end{figure}

\begin{figure}
  \begin{center}
  \includegraphics[width=0.8\textwidth]{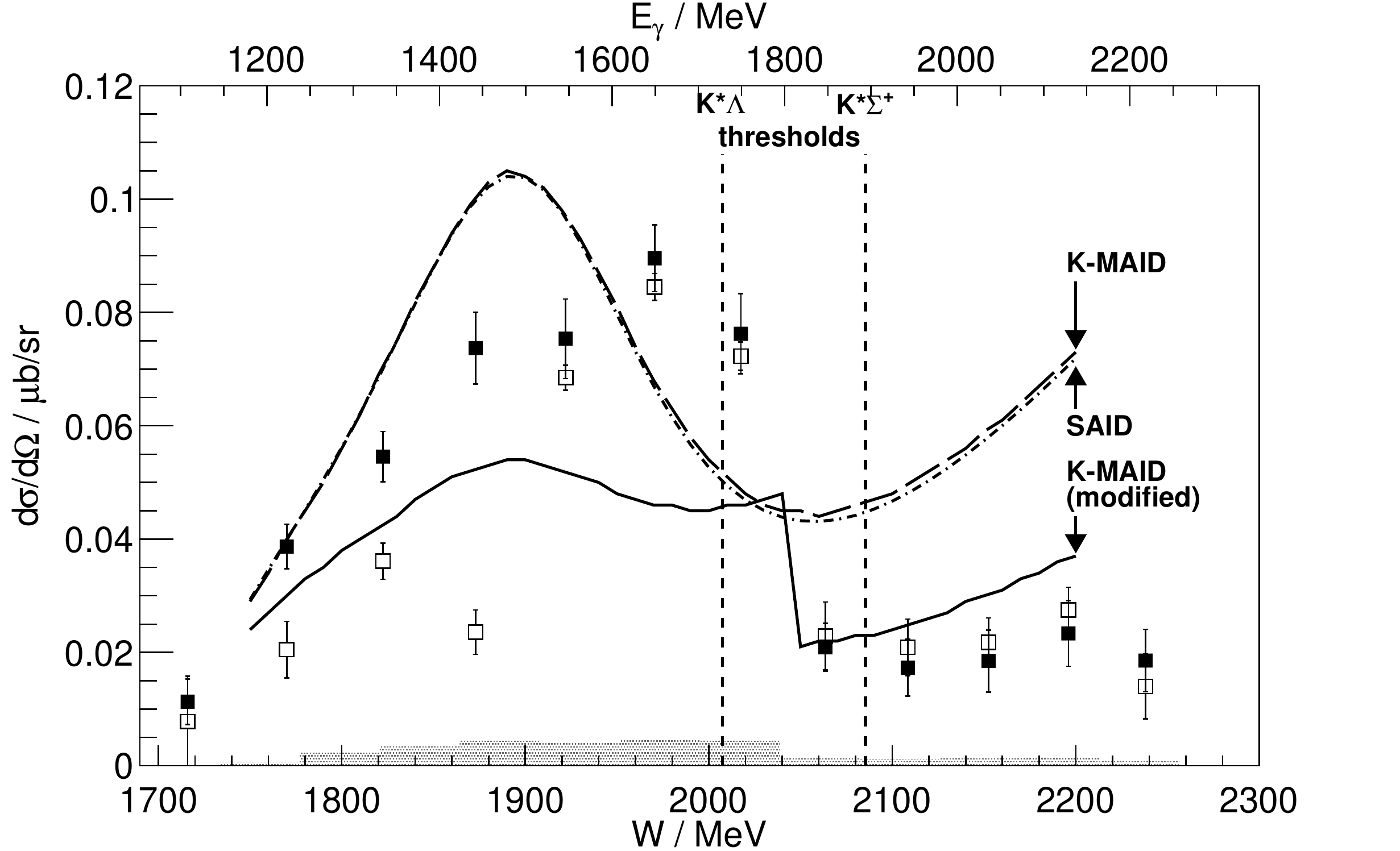}
  \end{center}
  \caption{\footnotesize %
Cross section for $K^0\Sigma^+$ photoproduction as a function of the 
centre-of-mass energy
from the present (full squares) and a previous (open squares) 
\cite{Castelijns08} 
Crystal Barrel experiment in the most forward angular bin of 
Figure\,\ref{fig:DiffWQ_comparison}. 
Plotted errors and curves represent the same as in 
Figure\,\ref{fig:total_x-sec},
the vertical lines as well.
         }
  \label{fig:forward_x-sec}
\end{figure}

As can be seen from Figure\,\ref{fig:DiffWQ_comparison},
directly above the $K^0\Sigma^+$ threshold a differential cross section
of $\simeq 0.02\,\mu$barn/sr is obtained with flat 
angular dependence, typical for s-wave production.
The cross section rises with increasing photon energy and also develops
an increasing forward peaking, suggesting increasing
$t$-channel contributions.
This forward peaking is most pronounced in the $E_\gamma = 1700 \pm 50$\,MeV
bin.
In sharp contrast, the next energy bin exhibits an entirely flat distribution 
again and a drop of the cross section back to $\simeq 0.02\,\mu$barn/sr.
This is at $E_\gamma = 1800$\,MeV, i.e. right between the thresholds of 
$K^*\Lambda$ and $K^*\Sigma$ photoproduction.
The differential cross section then remains almost flat and practically 
constant up to the highest measured energies.
This suggests that in the $K^*$ threshold region there is a sudden cross over
from a $t$-channel mechanism back to $s$-channel production of $K^0\Sigma^+$
with increasing photon energy.
As is shown in Figure\,\ref{fig:total_x-sec}, 
this effect is strong enough to become 
clearly visible in the total cross section 
which, given the full $4\pi$ acceptance, is simply obtained by 
integration of the differential cross sections.
Due to the forward peaking of the cross section below the $K^*$ thresholds,
the effect is most pronounced in the most forward angular bin, however. 
Here, the cross section drops by a factor of four in the vicinity of the 
$K^*$ thresholds, cf. Figure\,\ref{fig:forward_x-sec}.
It remains to be investigated whether a cusp-like structure develops, i.e. 
a discontinuity in the slope of the cross section. 

The structure seems related to a sudden change in the reaction mechanism.
Below the $K^*$ threshold, the angular distributions suggest that t-exchange
according to Figure\,\ref{fig:diagrams}(e) plays a major role, which is
associated with $K^{0*}$ exchange.
This hypothesis was tested with the K-MAID 
parameterisation \cite{K-MAID_webseite}, 
where it is possible to manually change the $K^*$ exchange strength.
With standard parameters both, K-MAID and SAID \cite{SAID}, 
deliver an unsatisfactory description of the data, 
cf. the dashed and dashed-dotted curves in Figure\,\ref{fig:total_x-sec},
respectively. 
Below the $K^*$ threshold this can be drastically improved by adjusting the 
couplings of the $S_{31}(1900)$ state to $G_1 = 0.3$ and $G_2 = 0.3$
\cite{K-MAID_webseite} and reduction of the Born-couplings from 1 to
$0.7$. 
To check whether the sudden drop in the total cross section can be 
reproduced by changing the $K^{0*}$ exchange, this contribution was
retained below the $K^*\Sigma^+$ threshold but manually set to zero above.
As is demonstrated by the full curve in Figure\,\ref{fig:total_x-sec},
with these modifications K-MAID yields
a significantly improved description of the cross section,
including the structure at the $K^*$ thresholds.
In forward directions (cf. Figure\,\ref{fig:forward_x-sec}) the modified
K-MAID still yields unsatisfactory results at smaller energies.
The drop of the cross section at the $K^*$ thresholds is however
rather well reproduced by switching off the $K^{0*}$ exchange contribution.

In the vicinity of the threshold, 
a $K^*$ would be produced almost on mass shell.
It then strongly couples to a $K^0$ and a pion.
We speculate that, in this way, 
close-to-threshold $K^*$ production feeds the $K^0$ channel:
The $K^0$ is ejected and observed in the final state while 
the pion is reabsorbed by the hyperon.
Such a process resembles diagram (f) of Figure\,\ref{fig:diagrams}.
A vector-meson - hyperon interaction in the intermediate state
is predominantly expected in a relative $s$-wave, 
which in turn would lead to the observed
flat $K^0$ angular distribution beyond the $K^*$ thresholds. 

If, above the $K^*$ threshold, the vector meson is produced as a
free particle, then diagram\,\ref{fig:diagrams}(f) no longer
contributes to the $K^0\Sigma^+$ channel.
The strength, which at the dip of the cross section is vanishing from  
the $K^0$ channel, is then expected to contribute to $K^{0*}\Sigma^+$.
In order to test this idea, the measured total cross section of the reaction
$\gamma p \rightarrow K^{0*}\Sigma^+$ \cite{Nanova08} was added to the
observed $K^0\Sigma^+$ cross section above the $K^*$ threshold.
The result is shown in Figure\,\ref{fig:total_x-sec} as open circles.
Using the sum of the two cross sections, a smooth transition is 
obtained from below to above the $K^*$ thresholds and the dip
structure vanishes.
This is taken as further indication of a production mechanism 
as outlined above.

This discussion suggests that
the situation depicted in Figure\,\ref{fig:diagrams}(f) 
can be seen as the coupling of the initial photon 
to a dynamically generated $(K^*\Sigma)^+$ or $(K^*\Lambda)^+$ 
state in the vicinity of the $K^*$ threshold.
The dip in the cross section occurs above the $K^*\Lambda$ but 
{\em below} the $K^*\Sigma$ threshold.
This fact may indicate that indeed an intermediate state is formed 
with a strong $K^*\Sigma$ component. 
Such states are expected in chiral unitary approaches
through the interaction of the nonet of vector mesons with 
the octet of baryons \cite{Oset-Ramos10}.
If the vector meson and baryon couple in a relative s-wave,
doublets of $J^P = (1/2)^-$ and $(3/2)^-$ are expected.
In ref.\,\cite{Oset-Ramos10} a non-strange isospin $1/2$ doublet 
is indeed predicted at a mass of 1972\,MeV, 
i.e. close to the $K^*$ threshold. 

It remains to be seen whether, in a partial wave analysis (PWA), 
the reported structures can be reproduced by $(1/2)^-$ and $(3/2)^-$
partial waves.
To make the PWA as unambiguous as possible, we are presently analysing
polarisation observables in addition to the cross sections.
In contrast to a $t$-channel dominated production mechanism,  
an $s$-channel intermediate state will provide a genuine
spin filter.
It is hence expected that, 
in addition to recoil polarisation and photon asymmetry, 
in particular the beam-target as well as the beam-recoil asymmetries will shed 
further light on the mechanism of $K^0$ photoproduction
in the vicinity of the observed dip structure. 

The reported structure in the cross section 
is also close to the $\eta' p$ threshold. 
In Ref. \cite{OR11} a significant coupling of vectormeson-baryon to
pseudoscalar-baryon channels with the same quantum numbers is expected.
Consequently, one may speculate that the possible $K^*$-hyperon states 
may affect the $\eta' p$ cross sections at threshold as well, and thus
help to solve the puzzle of $\eta' N$ interactions in both, 
hadronic and photoinduced reactions \cite{OR11}.

\section{Summary and outlook}

Using the Crystal-Barrel/TAPS detector setup at the electron accelerator 
facility ELSA of Bonn University, the reaction 
$\gamma + p \rightarrow K^0 + \Sigma^+$
was investigated from threshold to $E_\gamma = 2250$\,MeV.
We find an unexpected structure in the differential cross section 
between the $K^{0*}\Lambda$ and $K^{0*}\Sigma$ thresholds:
The angular distribution exhibits a sudden transition from forward peaked
to flat with increasing photon energy.
In forward directions the cross section drops by a factor of four,
generating a pronounced structure even in
the total cross section.
It is speculated that the effect may be due to close-to-threshold $K^{0*}$
production via the formation of a $K^*$-hyperon quasibound state,
as expected in Reference\,\cite{Oset-Ramos10}.
Above the $K^*$ threshold a real $K^*$ may be produced and the associated 
strength then vanishes from the $K^0$ channel. 
This is supported by the total cross section of $K^{0*}\Sigma^+$
photoproduction, which corresponds in size to the reduction of the 
$K^0\Sigma^+$ cross section in the dip region.

To shed more light on the threshold structure it will be mandatory
to exploit polarisation observables.
In particular, the photon beam asymmetry should be sensitive to the 
parity character of the $t$-channel contributions, while recoil polarisation
and beam-target asymmetry will strongly constrain the quantum numbers 
of an intermediate $s$-channel resonance.
In the meanwhile polarisation observables have been 
measured by the collaboration and are presently under analysis.
In addition, a partial wave analysis is under way where it will be 
interesting to see whether indications for $(1/2)^-$ and $(3/2)^-$ partial 
waves can be identified as they would be expected for a $K^*$-$\Lambda/\Sigma$
quasibound state.

\section*{Acknowledgements}

Helpful discussions with M. Lutz, E. Oset and A. Rusetsky 
are gratefully acknowledged.
We thank the staff and shift-students of the ELSA accelerator for
their enthusiasm to provide an excellent beam.
This work was supported by the federal state of 
{\em North-Rhine Westphalia} and the
{\em Deutsche Forschungsgemeinschaft} within the SFB/TR-16.
The Basel group acknowledges support from the
{\em Schweizerischer Nationalfonds},





\bibliographystyle{elsarticle-num}
\bibliography{<your-bib-database>}



\end{document}